\documentclass{article}
\usepackage{epsfig}
\usepackage{graphicx}

\title{VLBA Scientific Memorandum No. 32 \\
\vspace*{2.cm}
\underline{Multi-frequency ASTROMETRY with VSOP-2:} \\
\vspace*{0.5cm}
 An Application of \\
{\sc Source/Frequency Phase Referencing (sfpr)} techniques
\vspace*{2.cm}}

\author{
{\sc {\bf Mar\'{\i}a J. Rioja and Richard Dodson}} \\
University of Western Australia, UWA, Australia \\
Observatorio Astron\'omico Nacional, OAN, Spain} 

\begin{document}

\maketitle

\pagebreak


\begin{center}
\large
{\bf ABSTRACT} \\
\end{center}

\normalsize 

This document describes the advantages of applying {\sc
  source/frequency phase referencing} ({\sc sfpr})
techniques to the analysis of VLBI observations with VSOP-2, for high
precision astrometric measurements and/or increased sensitivity.  The
{\sc sfpr} calibration technique basics and a demonstration of the method
applied to highest frequency VLBA observations
are described in detail in VLBA Scientific Memo 31.\\
Here we outline its importance in the context of space VLBI
astrometry with VSOP-2,
where errors in the satellite orbit determination and
rapid tropospheric phase fluctuations 
set extreme challenges for the successful application of
{\sc conventional phase referencing} 
techniques, specially at the higher frequencies.
{\sc sfpr} is ideally suited for full calibration of those 
- regardless of the orbit determination accuracy - and, in
general, of any non-dispersive terms. 
The requirements for application of {\sc sfpr} techniques are fully 
compatible with current technical specifications of VSOP-2. Hence we
foresee that {\sc sfpr} will play an important role in helping
expanding the scientific outcome of the space VLBI mission. \\


\begin{center}
\large
{\bf The {\sc sfpr} technique} \\
\end{center}

\normalsize


\noindent
In this section we include a brief review of the basics of the {\sc
  source/frequency phase referencing} ({\it hereafter} {\sc sfpr})
technique, emphasizing the aspects which are relevant for space VLBI.
A full description is given in Dodson \& Rioja (2008, 2009). \\


This calibration approach relies on {\it fast frequency switching}
combined with {\it slow source switching} observations, between two
frequencies ($\nu^{high}$ and $\nu^{low}$) and two sources ({\it A}
and {\it B}), respectively.  The former provides extended coherence
time, and increased sensitivity, at the higher frequency; when
combined with the latter it gives astrometric capability to
observations with VSOP-2, irrespective of the uncertainties in the
orbit reconstruction which set an insurmountable limiting factor for
using {\sc conventional phase referencing} ({\it hereafter} {\sc PR})
techniques.

The {\sc sfpr} observations consist of alternating scans between the two
frequencies, with a duty cycle within the coherence time at the lower frequency.
The analysis is much simplified if the frequencies have an integer
ratio, {\it R = $\frac{\nu^{high}}{\nu^{low}}$}.


Following standard nomenclature, the residual visibility phases for
observations at the $\nu^{high}$ frequency of source $A$ 
are expressed as a compound
of geometric, 
tropospheric, ionospheric, instrumental and structural 
(non-zero for extended sources) terms:


\begin{equation}
\phi^{high}_{A} = \phi^{high}_{A,geo} +\phi^{high}_{A,tro} +\phi^{high}_{A1,ion}
+\phi^{high}_{A,inst} + \phi^{high}_{A,str} + 2\pi n
,\, \,\,\,\,\,\,n\,\,integer  
\end{equation}

\noindent
where $2\pi n$ stands for the phase ambiguity term.  The terms in
equation (1) result from the differences between the true
({\it labelled} ``true'') and {\it a priori} estimated ({\it labelled}
``mod'') values for the parameters used in the models involved in the data
analysis. For example, the geometric contribution 
accounts for: 

\begin{center}
 $\phi^{high}_{geo} = 2 \pi (\vec{D}_{\lambda}^{true}\,.\,\vec{s}^{true} -
\vec{D}_{\lambda}^{mod}\,.\,\vec{s}^{mod})$ \\
\end{center}

\noindent
where $\vec D_{\lambda}$ is the baseline vector in units of observed
wavelength, and $\vec s$ is the unit vector in the direction of the
source.  A simplified general expression for the tropospheric contribution
induced in an interferometer is:

\begin{center}
 $\phi^{high}_{tro} = 2 \pi \, \nu^{high} \,
(z_1 \,.\,f_1 - z_2 \,.\, f_2)/c $
\end{center}

\noindent
where $z_1$ and $z_2$ are the excess path length in the zenith
directions (e.g. $z = z^{true} - z^{mod}$), and $f_1$ and $f_2$ are
the mapping functions corresponding to
the source directions at the two antennas. 
For baselines between a ground and a satellite antenna, 
where the effect of the propagation 
medium needs to be considered only for the ground component, the 
expression above is simplified to include only one adding term.\\

A similar expression to equation (1) holds for the residual phases
$\phi^{low}_{A}$ from observations at $\nu^{low}$, the reference
frequency. These are analysed using self-calibration techniques. The
antenna based estimated corrections are linearly interpolated to the
times when the $\nu^{high}$ frequency is observed ($\tilde
\phi^{low}_{A,self-cal}$), scaled by the frequency ratio $R$
(with $R=\frac{\nu^{high}}{\nu^{low}}$), and applied as calibration of the
observed phases at $\nu^{high}$, the target frequency, in equation
(1).  We name this step as {\sc frequency phase transfer} ({\sc fpt}).
This calibration strategy results in perfect cancellation of the
non-dispersive rapid tropospheric residual phase fluctuation terms 
in equation (1), since:
\begin{center}
$\phi_{A,tro}^{high} - R\,.\,\tilde \phi_{A,tro}^{low} = 0 \, $ \\
\end{center}
\noindent
and the same applies for any orbit errors, and in general antenna and source 
coordinate errors contributing to the geometric term ($\phi_{geo}$) 
in observations of an achromatic source, since:
\begin{center}
$\phi^{high}_{A,geo}  - R\,.\,\tilde\phi^{low}_{A,geo}  =  0\, $ \\
\end{center}
\noindent
In general, for sources whose VLBI position is frequency dependent
the previous expression includes a ca. 24-hour sinusoidal extra term 
whose amplitude depends on the magnitude of the shift between the $\nu^{low}$
and $\nu^{high}$ observed frequencies, also known as ``core shift''
($\vec{\theta}_A$), and the baseline length:

\begin{center}
$\phi^{high}_{A,geo}  - R\,.\,\tilde\phi^{low}_{A,geo}  = 
2\pi \,\vec{D_{\lambda}}\,.\,\vec{\theta_{A}} + O (\vec{\Delta
D_{\lambda}}\,.\,\vec{\theta_A})$ 
\end{center}
\noindent
and also, it includes an extra contribution proportional to the scalar
product of the antenna position error (or orbit error) and the ``core
shift'' vectors, which we include in the formulation for
completeness. The effect of the extra term is negligable and can be
completely ignored given the VSOP2 orbit errors, or any other VLBI
antenna position errors, and the expected typical values for ``core
shifts''
for the range of frequencies observed with VSOP2. 

\newpage

\noindent
The resulting {\sc frequency phase transferred} ({\sc fpt}) 
visibility phases are: 

\begin{equation}
\phi_A^{FPT} = 
 \phi_{A,str}^{high} + 
2 \pi \, \vec{D_{\lambda}}\,.\, \vec{\theta}_{A} 
+ {\it \rm ``ion"}   
+  {\it \rm ``inst"}
\end{equation}
\noindent
 which contain, among others, contributions from the radio structure
 of the source at $\nu^{high}$, and, if present, the astrometric
 ``core shift'', which modulates each baseline with a ca. 24-hour
 sinusoid. We have omitted the $2\pi$ phase ambiguity term for
 simplicity, assuming that $R$ is an integer number. The rapid tropospheric
 fluctuations have been perfectly calibrated, but longer time scale,
 residual ionospheric and instrumental (dispersive) terms, named
 ``ion'' and ``instr'' remain. Middelberg et al. (2005) successfully
 applied the fast frequency switching technique to mm-wavelength VLBI
 observations as a means to extend the coherence time and allow
 detections of weaker sources.  They used one extra step of self
 calibration to get rid of the residual dispersive terms in equation
 (2). But this step of self-calibration also eliminates the position
 information, and hence prevented the measurement of the ``core
 shift''.

\noindent
An alternative approach to correct the residual long timescale phase drift
terms while preserving the astrometric information
consists of using the interleaved observations of a second source, a nearby
calibrator ($B$).
The angular separation and duty cycle are determined by the
ionospheric isoplanatic patch size, and the slow temporal properties 
of the  ``ion'' and ``instr'' terms in equation (2), respectively. 
For the case of interest here, the angular separation can be up
to several degrees, and the duty cycle of several minutes (ca. 5 minutes).
The observations of the calibrator source are first analysed using the same
strategy as described above for source ``$A$''. Next, they are 
used to correct the target source observations, in a similar fashion
as is done in conventional phase referencing observations of two sources
(using a temporal interpolation of antenna based solutions for {\it B} 
source, $\tilde \phi^{FPT}_{B,self-cal}$).
The residual {\sc source/frequency referenced} 
visibility phases ($\phi^{SFPR}_{A}$) are:

\begin{center}
$\phi^{SFPR}_{A} = \phi^{FPT}_{A} - \tilde \phi^{FPT}_{B,self-cal}$ 
$= \phi_{A,str}^{high} + 2 \pi \vec{D_{\lambda}} \, . \, (\vec{\theta}_{A}-
\vec{\theta_{B}})$ 
\end{center}

\noindent
The combination of the frequency and source switching in {\sc sfpr} 
results in calibrated phases free of long scale drift terms.
\noindent
Finally, a Fourier transformation of the phases, without further
calibration, results in a map of the brightness distribution of the
target source ($A$) at $\nu^{high}$ frequency, and where the offset
of the peak with respect to the center of the map is astrometrically
significant: a measurement of the relative ``core shift'' between both
frequencies ($\nu^{high}$ and $\nu^{low}$), for both sources ({\it A} and
{\it B}).

\begin{center}
\large
{\bf  {\sc sfpr} observations with VSOP-2}
\end{center}


\normalsize 


\noindent
VSOP-2 is the second space VLBI mission
planned by the Institute of Space and Astronautical Science (ISAS). 
Its goal is placing a satellite radio telescope in orbit for joint 
VLBI observations with ground radio telescopes.
It will be equipped with a 9.3 meter off-axis parabolic antenna, and
dual polarization receivers at 8.4, 22, and
43-GHz. \\

%
\noindent
The analysis of data from its predecessor VSOP mission,
with the HALCA satellite, showed that the calibration of space 
VLBI observations
involves additional 
difficulties arising from the relatively poor sensitivity achievable
with small orbiting antennas, smaller correlated source flux densities
at the higher resolution of space baselines, and large geometric 
delay errors introduced by uncertainties in the
spacecraft orbit that prevent long integration times (Rioja \& Porcas,
2007).
Additionally, VSOP-2 is designed to carry out astrometric observations,
which impose more severe demands on the phase calibration. \\



\noindent
The use of conventional {\sc pr} calibration techniques for
space VLBI data analysis, especially at the higher frequencies, will be
rather challenging, due to the shortage of suitable reference sources
(i.e. compact and strong enough on space baselines), rapid tropospheric phase
fluctuations and satellite orbit determination errors.
Asaki and collaborators (2007) have carried out a comprehensive study
on the feasibility of conventional {\sc pr} observations with VSOP-2,
under a range of different weather
conditions and orbit determination accuracy of the satellite, among
other parameters. Their simulations demonstrate that astrometrical observations
are expected to achieve a good performance at 8.4 GHz, while at the 
higher frequencies the best possible weather would be required and the 
probability of finding suitable calibrators (particularly at 43 GHz)
is greatly reduced.\\

\noindent
Alternatively, VSOP-2 could benefit from the use of the {\sc sfpr} technique for
astrometric studies and/or increasing coherence time at the higher
frequencies. It encompasses observations at two frequencies, ideally
with an integer ratio, and of 
two sources. 
The VSOP-2 mission specifications are compatible with
the observing schedule requirements for successful application of our
{\sc sfpr} calibration method, that is, to have the capability to alternate
observations between frequency bands, e.g. between 21.5 and 43 GHz,
with a ca. 1 minute cycle ({\it fast frequency switching}), and
between sources with a much longer cycle, of several minutes
({\it slow source switching}).\\

\noindent
The {\sc sfpr} route holds great promise for space VLBI since any orbit
errors (as well as ground antennae coordinate errors) are fully
removed in the analysis, therefore alleviating the constraint on orbit
determination accuracy; also, the conditions for a suitable {\sc sfpr}
calibrator source are much more relaxed than in conventional {\sc pr},
and the angular separation between sources can be up
to several degrees, and the observing source duty 
cycle up to several minutes.\\

\noindent
Figures 1 to 3 show the estimated
error budget for astrometric analysis of VSOP-2 observations
using conventional {\sc pr}, and {\sc sfpr} techniques, for comparison.
The individual mis-modelled contributions are shown in separate 
columns, and labelled following standard nomenclature: 
uncertainties arising from satellite orbit error determination ({\sc
  odda}), errors in coordinates of reference source (RefSource),
unmodelled propagation effects through the troposphere ({\sc trop})
and ionosphere ({\sc ion}), and finally the limitation imposed by the
finite interferometric resolution and noise in the observables ({\sc
  resol-snr}).\\

\noindent
Figure 1 shows the astrometric error budget for conventional {\sc pr}
observations at 8.4 GHz, of a pair of sources $2^o$ apart with a
one-minute duty cycle, with a space baseline and different orbit
errors. 
The plot shows the ODDA contributions arising from increasing orbit
errors of 3, 5 and 10 cm; we used 
``typical values'' for the parameter model errors, as listed in Asaki
et al. (2007), to estimate the rest of the contributions.
The current plan for precise orbit determination for ASTRO-G, 
at the time of writing this memo, is described in Asaki et al. (2008).
Residual tropospheric/ionospheric errors are
the dominant contributions with orbit errors less than 5 cm, at 8.4
GHz.  All residuals scale with the pair angular separation, except for
the thermal noise contribution.  Also, since all residuals scale with
baseline length (except for the ``RefSource'' contribution) the
long space baselines result in an
improvement with respect to a ground array. \\
 

\begin{figure}
\epsfig{file=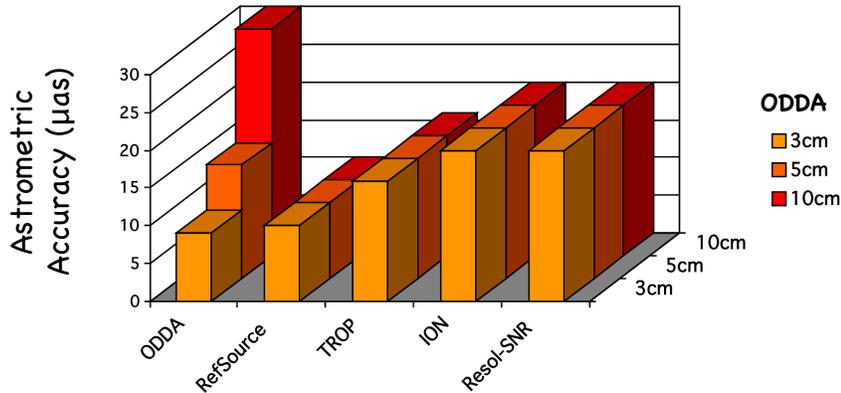,angle=90,width=11cm} 
\caption{{\it Error budget for \underline{{\sc conventional pr}-astrometric
    analysis} of space VLBI observations at 8.4 GHz of a pair of
    sources $2^o$ apart, for increasing
    orbit errors (shown with different colours). The columns account
    for individual contributions arising from satellite orbit errors
    (ODDA), reference source coordinates, tropospheric and ionospheric
    residuals and the instrumental resolution and SNR criteria. Orbit 
    errors equal to 3, 5, and 10 cm, and ``typical values''
    for the rest of the parameter model errors (see text).
    Tropospheric/Ionospheric terms are among the dominant ones with
    orbit errors less than 5 cm, at 8.4 GHz.}}
\end{figure}

\noindent
Figure 2 shows the astrometric error budget for conventional {\sc pr} 
observations at 8.4, 22
and 43 GHz, with an orbit error of 3cm - all other parameters are as for
Figure 1. Contributions from non-dispersive terms, such as geometric
contributions (e.g. orbit and reference source coordinate errors)
remain identical at the 3 frequency bands.
Instead, the ionospheric term is dispersive, e.g. decreases with
increasing frequency, and is much reduced at 22 and 43 GHz, with respect to
8.4 GHz observations.\\

\begin{figure}[htbp]
\vspace*{-1.5cm}
\epsfig{file=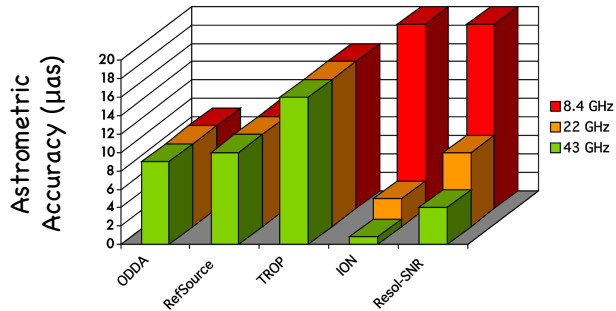,angle=90,width=8cm} 
\caption{{\it Error budget for \underline{{\sc conventional pr-}astrometric
    analysis} of space VLBI observations at 8.4, 22 and 43 GHz,
    assuming an orbit error of 3cm. 
    Contribution from non-dispersive terms remains identical
    irrespective of the observing band $\sim 20 \mu$as. The ionospheric
    term is dispersive, and is much reduced at 22 and 43 GHz.}}
\end{figure}




\noindent
Figure~3 shows a compilation of {\sc sfpr-}astrometric error budget
estimates for observations of eligible frequency pairs observed with
VSOP-2, 
with a one-minute frequency switching cycle.   The
first striking aspect is the perfect compensation of the non-dispersive
terms, that is, the orbit errors are cancelled out in the
analysis.

\begin{figure}[Hb]
\epsfig{file=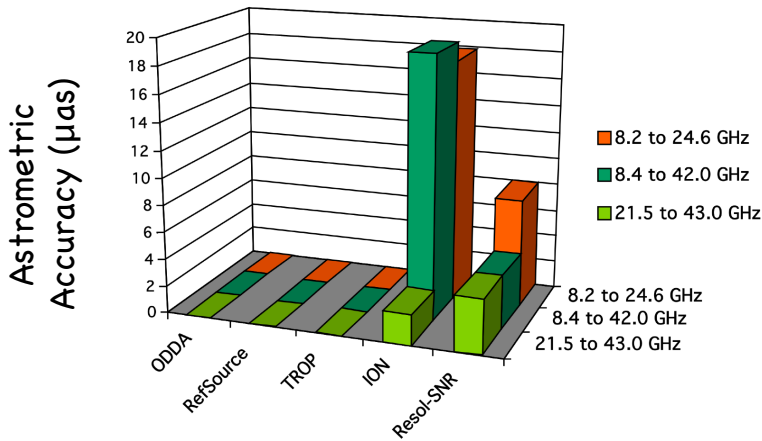,angle=90,width=10cm} 
\caption{{\it Error budget estimates for \underline{{\sc sfpr-}astrometric
    analysis} of space VLBI observations, switching between pairs of frequencies
    in VSOP-2 with a duty cycle of 60 seconds.  Parameter model errors as in
    Figures 1 and 2.
    Note the perfect compensation of non dispersive
    terms. The residual ionospheric contribution is approximately that
    from the the reference (lower) frequency scaled by the frequency
    ratio.The ionospheric contribution is the dominant one when 8 GHz is the
    reference frequency. The improvement becomes noticeable in the
    combination 22 to 43 GHz, when the 22 GHz ionospheric contribution is
    very much reduced.}}
\end{figure}

 The residual ionospheric contribution depends on the
reference frequency ($\nu^{low}$), and the frequency ratio ($R$); 
the higher the reference
frequency  and the smaller the frequency ratio, the smaller
the contribution to the error budget. The mis-modelled ionospheric
contribution is the dominant source of error when using 8.4 GHz as the
reference frequency. The improvement becomes noticeable between 22 to
43 GHz, when the residual ionospheric contribution is very much reduced 
at 22 GHz. \\

\begin{center}
\large
{\bf Summary}  \\
\end{center}

\normalsize

The current technical specifications of VSOP-2 are compatible with
the requirements to carry out successful {\sc sfpr} observations in order to
achieve astrometric measurements and/or increase the coherence time in
VLBI observations with the space mission.
Moreover, the {\sc sfpr}
technique naturally addresses the main factors which limit the
application of conventional phase referencing to VSOP-2 observations,
i.e.
orbit uncertainties and tropospheric fluctuations, specially
at the higher frequencies. A list of benefits from applying {\sc sfpr}
to VSOP-2 space mission follows:

\begin{itemize}
\item {\sc sfpr} technique relies on fast switching between
  frequencies, rather than
between sources as in conventional phase referencing, 
a less demanding operation for VSOP-2. 
\item Satellite Orbit uncertainties, which set a hard constraint for using
conventional phase referencing, are not
an issue for the {\sc sfpr} technique. 
\item  A larger number of
calibrators available for {\sc sfpr} since wider angular separation is 
acceptable.
\item {\sc sfpr} techniques result in
increased sensitivity at 43 GHz, and recoverable astrometric information 
regardless of the orbit determination accuracy.
\item The implementation of {\sc sfpr} techniques are compatible with
  VSOP-2 technical specifications.
\end{itemize}

Summarizing, we think this new technique will prove its potential in
the calibration of observations at the highest frequency, 43 GHz,
either using the second harmonic of 21.5 GHz, or the fifth harmonic of
8.4 GHz as ``reference'' observations. The current mission specifications
do not allow the calibration of the 22 GHz band
with the 8.4 GHz band, unless the latter is expanded to cover 7.5 GHz.\\

\noindent
\large
{\bf Acknowledgements} \\

\normalsize


The authors express their gratitude to Craig Walker, Ed Fomalont, Asaki
Yoshiharu and Richard Porcas for providing useful comments on this manuscript.

{}


\begin{thebibliography}{}

\bibitem{}Asaki, Y., Sudou, H., Kono, Y., Doi, A., Dodson, R., Pradel, N.,
Murata, Y., Mochizuki, N., Edwards, P., Sasao, T., and Fomalont,
E. 2007. \  Verification of the Effectiveness of VSOP-2 Phase
Referencing with a Newly Developed Simulation Tool, ARIS.\ PASJ 59,
397. \\

\bibitem{} Asaki, Y., Takeuchi, H., Yoshikawa, M., 2008, Next
  Space-VLBI Mission, VSOP-2, and the Precise Orbit Determination with
  GNSS Navigation and SLR, Proc. of ION GNSS 2008 \\

\bibitem{} Dodson, R. \& Rioja, M. 2008, Informe T\'ecnico, IT-OAN-2008-03 \\

\bibitem{} Dodson, R. \& Rioja, M. 2009, VLBA Scientific Memorandum No. 31 \\

\bibitem{} Middelberg, E., Roy, A.~L., Walker, R.~C., Falcke, H.\
  2005.\ VLBI observations of weak sources using fast frequency
  switching.\ Astronomy and Astrophysics 433, 897-909. \\

\bibitem{} Rioja, M. \& Porcas, R. \ 2007.\ Astrometry with VSOP.\ 
``Approaching Micro-Arcsecond Resolution
with VSOP-2: Astrophysics and Technology'', 
Eds. Y.Hagiwara, E.Fomalont, M. Tsuboi, and Y.Murata
(ASP Conference Series vol 402, 2009).
\\

\end{thebibliography}
\end{document}